\documentclass[12pt,preprint]{aastex}
\usepackage{graphics}
\begin{document}

\title{On Tripolar Magnetic Reconnection and Coronal Heating}
\author{${Kumud Pandey}^1$, ${Udit Narain}^1$ and ${N.K.Lohani}^2$}
\affil{1. Astrophysics Research Group,Meerut College, Meerut, India-250001}
\affil{2. Department of Physics, M.B.Govt.P.G.College, Haldwani, Nainital, India-263141}
\email{1. narain@iucaa.ernet.in, 2. lohani@iucaa.ernet.in}

\begin{abstract}

Using recent data for the photosphere-chromosphere region of the solar 
atmosphere the magnetic reconnection in tripolar geometry has been 
investigated through the procedure of Sturrock (1999). Particular 
attention has been given to the width of the reconnecting region, 
wave number of the rapidly growing tearing mode, island length scales, 
frequency of MHD fluctuations, tearing mode growth rate, energy 
dissipation rate and minimum magnetic field strength required to 
heat chromospheric plasma to coronal temperatures. It is found that 
small length scales are formed in the upper chromosphere. The maximum growth 
rate of tearing mode instability coincides with the peak in the 
energy dissipation rate both of which occur in the upper chromosphere 
at the same height. It is realized that the distribution of magnetic field  
with height is essential for a better understanding of the coronal heating 
problem.\\  

Subject headings: Sun: chromosphere - Sun: corona - Sun: magnetic field

\end{abstract}

\section{INTRODUCTION}

Recently \citep{asc01} has used Yohkoh, Solar Heliospheric Observatory (SOHO) and Transition Region And Coronal Explorer (TRACE) observations in the 
evaluation of coronal heating models for active regions. He focusses on three 
main results obtained from aforesaid spacecrafts, namely, the overdensity of 
coronal loops, chromospheric upflows of heated plasma and the localization 
of heating in the lower corona. He examines various A.C. and D.C. models on 
the basis of above criteria. In D.C. models he considers bipolar, tripolar 
and quadrupolar magnetic reconnection and favours tripolar magnetic 
reconnection where magnetic reconnection between two polarities of a closed 
and the unipolar footpoint of an open field line may take place. Such a 
situation may arise due the emergence of a new dipole in an open field region 
or by the collision of the footpoint of an open field line with an adjacent 
dipole. In this case the open field line can carry hot plasma upward into the 
upper corona, regardless of the size of the interacting dipole. Thus tripolar 
reconnection is highly relevant for the coronal heating especially for open 
field regions such as coronal holes.\\

\citep{cha00} have studied transient network brightenings (blinkers) 
and explosive events in the solar transition region, recorded by the Coronal 
Diagnostic Spectrometer (CDS) in the line O V $\lambda$ 630 and Solar 
Ultraviolet Measurements of Emitted Radiation (SUMER) instrument in the line Si IV 
$\lambda$ 1402 together with the photospheric magnetograms taken by the Big 
Bear Solar Observatory videomagnetograph. They find that the explosive events 
(which are features with very broad UV line profile) tend to keep away from 
the centres of network brightenings. CDS  blinkers contain many small-scale 
short-lived SUMER `unit brightening events' having size of a few arc seconds 
and a lifetime of a few minutes. Each of these unit brightening events is 
characterized by a UV line profile which is not as broad as those of explosive 
events. Thus blinkers (transient network brightenings) and explosive events
both may be due to magnetic reconnection with different geometries.\\

The explosive events are considered to occur as a result of the collision of a 
network flux thread (which is part of a very large loop) and an intranetwork 
flux thread which is part of a very small loop ( \citet{cha99}, 
\citet{cha00} ). This interaction has therefore the potential for conveying hot plasma into large-
scale loops and is a very suitable mechanism for explaining the observed footpoint 
heating, upflows and overdensity in EUV loops. Further, the relaxation of the 
newly formed open field line may accelerate acoustic waves or shock waves and 
heat plasma along its passage. Thus the process of tripolar magnetic reconnection 
has a better connectivity to the upper corona and to the solar wind via the open 
field line than bipolar and quadrupolar reconnection geometries \citep{asc01}.\\ 

When two open magnetic field lines of opposite polarity reconnect, the newly 
configured polarities relax into two bipolar field lines. The lower one connects 
the two magnetically conjugate photospheric footpoints whereas the upper ends 
of the disconnected field lines in the upper corona combine and finally move into 
interplanetary space. Thus there is  a high-density closed loop downward and a 
low-density segment moving upward (see, e.g., \citet{shi92}; \citep{kru00}; \citep{par00}; \citet{asc00} ). This type of heating process is applicable to big 
or small flares (including micro- and nano-flares) (\citep{par91}; \citep{bro00}). 
Bipolar magnetic reconnection has been investigated, in some way or other, by 
\citep{lit99} , \citep{lon99} , \citep{fur00} , \citet{sak00} , \citep{sak01} also.\\

Quadrupolar magnetic reconnection has been observed and modeled in solar flares 
( \citet{uch94}; \citep{ura96}; \citep{han96} ; \citet{nis97}; \citet{asc99}). Here the exchange of connectivity between positive and negative polarities 
in a system with two closed field lines takes place by pushing them into physical 
contact such that the new configuration is in a lower energy state.\\

Chromospheric quadrupolar magnetic reconnection between two contiguous flux tubes 
of opposite polarity has been considered by \citep{stu99} in connection with its 
possible relationship to coronal heating. His basic idea has been that reconnection 
occurs preferentially where the growth rate of the relevant instability is greatest. 
He arrives at the conclusion that quadrupolar magnetic reconnection can lead to 
coronal heating by Joule dissipation and by the generation and subsequent dissipation 
of  high frequency Alfv\'en and magnetoacoustic waves. The heating of solar wind 
particles may take place when high frequency Alfv\'en waves, produced during 
reconnection, are absorbed by cyclotron damping.\\

In view of above it seems quite worthwhile to study magnetic reconnection in 
tripolar geometry such as that given in Fig. 1. In \S  2 we present the 
relevant theoretical details. The data used and the results obtained, exhibited 
in \S  3, are discussed in \S 4 alongwith our conclusions.\\

Throughout cgs system of units is used. Wherever necessary, the heights are expressed 
in km.

\section{THEORETICAL DETAILS}

We adopt the formalism of \citep{stu99}. As shown in Fig. 1 there is a 
network flux thread (part of a very large loop) which collides with an intranetwork 
flux thread which is a part of very small loop. It is assumed that magnetic 
reconnection would occur preferentially where the growth rate of resistive tearing 
mode instability is greatest. The medium under consideration is a partially ionized 
hydrogen plasma so that the role of helium and other elements may be ignored.\\

The tearing mode growth rate depends on the wave number of the mode. The wave number 
of the most rapidly growing mode is given by \citep{stu94}

\begin{equation}
 k_{M}  = 1.4 R_{m}^{-0.25} l^{-1},
\label{eq1}
\end{equation}

where $l$ is the width of the reconnecting region and $R_{m}$, the transverse magnetic 
Reynolds number, is given by 
\begin{equation}
R_{m} = t_{D} / t_{A}, 
\label{eq2}
\end{equation}

with 

\begin{equation}
t_{D} = 4{\pi} l^2 / {\eta} c^2,
\label{eq3}
\end{equation}

and 

\begin{equation}
t_{A} = l / v_{A} .
\label{eq4}
\end{equation}

Here $t_{D}$ is the resistive diffusion time, $t_{A}$ the Alfv\'en transit time, $v_{A}$ 
the Alfv\'en velocity, c the speed of light in vacuum and $\eta$, the Coulomb 
resistivity of the hydrogenic plasma, is given by \citep{spi62} 

\begin{equation}
\eta = 1.5  10^{-7} T^{-1.5}    s
\label{eq5}
\end{equation}

In equation (5),  T is the temperature of the hydrogenic plasma. The Alfv\'en 
velocity $v_{A}$ is estimated through the following equation

\begin{equation}
v_{A} = B / ( 4 \pi \rho )^{0.5},
\label{eq6}
\end{equation}

where the mass density $\rho$ may be obtained from
\begin{equation}
\rho = (n_{p} + n_{H} ) m_{p}. 
\label{eq7}
\end{equation}

In the above B is the magnetic field strength before reconnection, $n_{p}$ and $n_{H}$ 
are the number densities of proton and neutral hydrogen, respectively.\\

The growth time $t_{g}$, which is the inverse of the growth rate, is given by

\begin{equation}
t_{g} = 1.585 R_{m}^{-0.5} t_{D}.
\label{eq8}
\end{equation}

The growth rate of instability depends sensitively upon the assumed width $l$
 of the 
current sheet between two flux  tubes and it may be related to the pressure scale 
height H by
\begin{equation}
l = {\epsilon} H.
\label{eq9}
\end{equation}

Here $\epsilon$ is a dimensionless parameter and the scale height $H$ is given by
\begin{equation}
H = 10^{3.48} T m_{p} / m_{av},
\label{eq10}
\end{equation}

with 
\begin{equation}
m_{av} = (n_{p} + n_{H}) m_{p} / (n_{e} + n_{p} + n_{H}), 
\label{eq11}
\end{equation}

where $m_p$ is the proton mass and $n_e$ the number density of electrons.

The tearing mode instability leads to the formation of small islands which allow 
for the rapid diffusion of the magnetic field. For the most rapidly growing mode 
the island length scale $x_{TM}$ is given by
\begin{equation}
x_{TM} = 1.51 R_{m}^{-0.25} l.
\label{eq12}
\end{equation}

If the tearing region is in a dynamic state and the inhomogeneities move at the 
Alfv\'en speed then they will lead to fluctuations having frequency $\nu_{TM}$ 
given by 
\begin{equation}
\nu_{TM} = v_{A} / x_{TM}.
\label{eq13}
\end{equation}

Such fluctuations may play a role in providing additional thermal energy to 
the corona.\\

The average energy dissipation rate in the reconnection region may be obtained from(Tandberg-Hanssen \& Emslie 1988)
\begin{equation}
U_{t} = B^2 V / 8 {\pi} t_{R},
\label{eq14}
\end{equation}

where $V$ is the volume of the reconnecting region(s) and $t_{R}$, the reconnection  time scale, is given by \citep{spi77} 
\begin{equation}
t_{R} = t_{A}^{0.4} t_{D}^{0.6}. 
\label{eq15}
\end{equation}

In the case of multiple reconnecting regions, as in the tearing mode, the 
volume of reconnecting region may be taken to be $V = L^2 l$, where L is 
the longitudinal length scale and $l$ is the width. As an approximation L 
may be taken to be the diameter of the flux tube.\\

It is possible to estimate the minimum strength of the magnetic field required 
to heat the reconnecting region to coronal temperature ($10^6$K) by using the 
following relation
\begin{equation}
1.5 n k_{B} T = B_{min}^2 / 8 {\pi},
\label{eq16}
\end{equation}

where $k_{B}$ is the Boltzmann constant and n is given by
\begin{equation}
n = n_{e} + n_{p} + n_{H}.
\label{eq17}
\end{equation}

Such an estimate can be made at each height where magnetic reconnection 
can occur.

\section{THE DATA AND RESULTS}

For the sake of completeness the data taken from Cox (2000) have been 
exhibited in Figures 2 and 3. In particular, Fig. 2 shows variation of 
temperature with height in the photosphere-chromosphere region whereas 
Fig. 3 exhibits $n_{e}$, $ n_{p}$ and $n_{H}$ as a function of height h in km.



The data of Fig. 2 and equation (5) have been used to obtain resistivity 
$\eta$ as a function of height and is displayed in Fig. 4. Since the 
resistivity is inversely proportional to $T^{1.5}$ it has maximum value in 
the temperature minimum region.

The data of Figures 2 and 3 together with equations (\ref{eq10}) and ( \ref{eq11}) lead to the results displayed in Fig. 5 where the scale height H is shown 
to vary with height in the lower solar atmosphere. The lowest value of scale 
height falls in the temperature minimum region.\\

In order to estimate the width $l$ of the reconnecting region we require the 
value of parameter $\epsilon$. This can be done by estimating the length over 
which the magnetic field will diffuse over a time scale characteristic of 
changes in the photosphere, e.g., the mean lifetime of photospheric granules 
which is 10 minutes (\citep{cox00}). This procedure alongwith equations (\ref{eq3}), (\ref{eq5}), (\ref{eq9}) and the data of Fig. 2 give $\epsilon = 0.0114$ 
corresponding to $l = 1.48$ km and $T = 4400 K$ at 520 km. Now equation (\ref 
{eq9}) alongwith the results of Fig. 5 may be used to obtain the width $l$ of  
the reconnecting region as a function of height which is displayed in Fig. 6.



The nature of variation of width with height is similar to that of scale height  vs height curve, as expected.

The evaluation of Alfv\'en transit time $t_{A}$ requires the values of Alfv\'en 
velocity $v_{A}$ and the width $l$. In order to determine $v_{A}$ we require the 
magnetic field strength as a function of height (which is not known) and the 
mass density $\rho$. Assuming a constant mean value $B = 100$ G, the data of 
Fig. 3 and equation (\ref{eq7}), the Alfv\'en velocity has been determined and 
is displayed in Fig. 7 as a function of height. 

As expected $v_{A}$ increases with height because $\rho$ decreases with height. Using the values of $l$ and $v_{A}$ the Alfv\'en transit time becomes known 
through equation (\ref{eq4}).\\

With the width $l$ and the resistivity $\eta$ known it is possible to 
determine resistive diffusion time $t_{D}$ by using equation (\ref{eq3}). This 
enables us to find transverse magnetic Reynolds number $R_{m}$ through equation 
(\ref{eq2}). The Reynolds magnetic number has been displayed in Fig. 8 as a 
function of height. The resulting curve shows a minimum near 100 km and 
increases upwards monotonically. The resistive diffusion time $t_{D}$ is 
displayed in Fig. 9 as a function of height.


Using equation (\ref{eq1}), the values of $R_m$ and width $l$ it is possible 
to evaluate the wave number $k_M$ of the most rapidly growing mode. This wave 
number as a function of height is displayed in Fig. 10. It shows a broad 
maximum in the 0 - 700 km region.\\ 

The growth time (which is the inverse of growth rate) has been obtained through
Equation (\ref{eq8}) and the already known values of the magnetic Reynolds 
number (Fig. 8) and the resistive diffusion time (Fig. 9). The growth time 
$t_{g}$ and the growth rate are displayed in Figures 11 and 12, respectively.


The growth rate curve shows a maximum at 1800 km and the growth time curve 
shows a minimum at the same height, as it should.

Using Equation (\ref{eq15}), the resistive diffusion time t$_{D}$ and the 
Alfv\'en transit time t$_{A}$ we obtained reconnection time t$_{R}$ which is 
exhibited as a function of height in Fig. 13. It exhibits a broad minimum 
in the range 1200 - 1900 km. It appears that the reconnection is slower in 
the temperature region and faster in the upper chromosphere.\\ 

Fig. 14 exhibits island length scales $x_{TM}$ as a function of height. These 
values have been obtained from Equation (\ref{eq12}), the magnetic Reynolds 
number (Fig. 8) and the width $l$ of the reconnecting region (Fig. 6).  

The shortest length scales are obtained around 1900 km in the upper chromosphere which is a desirable feature for all coronal heating mechanisms \citep{asc01}.\\
The frequencies of MHD waves generated as a result of fluctuations in the 
reconnecting region are displayed in Fig. 15. These frequencies increase from 
temperature minimum region to upper chromosphere, similar to \citep{stu99}, 
monotonically.

Using Equation (\ref{eq14}), $B = 100$ G, $L = 10^8$ cm, the width $l$ 
(Fig. 6) of reconnecting region and reconnection time $t_{R}$ (Fig. 13) we 
estimate the average energy dissipation rate as a function of height. This 
result is displayed in Fig. 16 which shows that the energy dissipation rate 
increases with height upto 1800 km and falls upwards. 
. The peak in dissipation rate agrees with the peak in the growth rate of the tearing mode instability. The height of peak dissipation rate finds favour 
with the observation made by \citep{asc01}.

In Fig. 17 we display the minimum magnetic field strength $B_{min}$ required 
to heat the reconnecting site to coronal temperatue $10^6 K$ as a function 
of height. It is clear from this  curve that at a height of 1600 km the 
required minimum magnetic field is about 100 G whereas at a height of 500 km 
it is about 6000 G. Such a possibility is consistent with the evidence for direct heating of chromospheric gas to coronal temperatures \citep{kru98}.

\section{DISCUSSION AND CONCLUSIONS}

The data, namely, temperature $T$, electron, proton and hydrogen number 
densities $n_{e}, n_{p}, n_{H}$ and mean lifetime of granules, used by us are of 
quite recent origin and seem to be reliable. The temperature minimum occurs 
near 500 km (Fig. 1). At each height the number density of neutral hydrogen 
is greater than the number densities of electrons and protons (Fig. 3).\\

The classical Coulomb resistivity $\eta$ is inversely proportional to $T^{1.5}$ 
consequently it has largest value at that height at which the temperature 
is minimum. Similar trend is noticeable at other heights (Fig. 4).\\

The scale height $H$ and width $l$ of the reconnecting region vary with 
height in the same way. Both of them have their minimum values near 500 km. 
Essentially, the nature of curves of temperature, scale height, width, 
resistivity  is quite similar, as expected.\\

Because of decreasing mass density $\rho$ the Alfv\'en velocity increases 
with height, monotonically as the magnetic field strength $B$ before 
reconnection has been given a fixed value of 100 G. For the better 
understanding of the coronal heating problem the distribution of magnetic 
field with height is quite essential.

The magnetic Reynolds number determines to what extent the field is frozen 
into the plasma. Higher values of $R_{m}$ means better frozen-in-field 
approximation. In the region under consideration $R_{m}$ varies from about 
$3 \times 10^3$ to $3 \times 10^7$, i.e. the frozen-in-field approximation is quite 
good to describe photospheric-chromospheric region (Fig. 8).\\

The resistive diffusion time $t_{D}$ varies with height (Fig.9) in the same way as 
the temperature. It has minimum value near 500 km and lies in the range 
$6 \times 10^2 s < t_{D} < 6 \times 10^4 s$ for the solar atmospheric region under 
consideration.\\

The wave number of the rapidly growing mode shows a broad peak in the region 
0 - 700 km (Fig. 10). This is because $k_{M}$ is inversely proportional to 
the width of reconnecting region which has minimum value in the aforesaid 
region.\\

The growth time of instability decreases monotonically with height upto 
1900 km and increases upwards (Fig.11). Since the growth rate is the 
inverse of growth time  hence growth rate exhibits opposite nature (Fig. 12). 
The growth rate is maximum at 1800 km consequently heating is expected 
to occur near this height. In fact, the energy dissipation rate is maximum 
at 1800 km (Fig. 16). Thus the growth rate seems to be a better indicator 
of heating. Further the magnetic reconnection time $t_{R}$ has minimum value 
in the 1200 - 1900 km region (Fig.13) but the width of the reconnecting 
region is larger in this region. Since the energy dissipation rate varies  
directly with  width and inversely with reconnection time hence the above-
mentioned behaviour of dissipation rate is quite expected.\\

It is clear from Fig. 14 that the island length scale has its lowest value 
at 1900 km. Since smaller length scales are more efficient in energy 
dissipation than larger length scales hence the above heating rate is 
in agreement with this well-known result.\\

The energy dissipation rate (Fig. 16) gets support from the minimum magnetic 
field strength vs height curve (Fig. 17) because at 1900 km the required 
minimum magnetic field strength $B_{min}$ to heat the chromospheric plasma to 
coronal temperature is about 100 G. It is this value which has been used to 
obtain Alfv\'en velocity $v_{A}$, Alfv\'en transit time $t_{A}$ etc. .\\

The frequency of MHD waves, generated due to magnetic reconnection, increases 
with height (Fig. 15). Since the higher frequency waves may dissipate more 
readily than the lower frequency waves hence the heating wil be enhanced by 
the high frequency waves higher up in the atmosphere. Similar conclusion 
has already been forwarded by \citep{stu99}.

In tripolar geometry (Fig. 1) a network flux thread which is a part of a 
very large loop collides with an intranetwork flux thread which is part of  
a very small loop. At the intersection point the two flux threads are 
antiparallel, forming a relative angle that is greater than 90 degree. 
Consequently the reconnection at the intersection point could produce 
strong bidirectional outflows (Fig. 1a). This explains explosive events 
showing high velocity motions in their line profiles.

The open field line may carry hot plasm upward into the upper corona, 
regardless of the size of the interacting dipole. The open field line 
may also act as a waveguide to send MHD fluctuations into the corona and 
interplanetary space. 

Thus tripolar reconnection is a very suitable mechanism for heating open 
field regions (coronal holes) in particular and the solar corona in general.
It may also accelerate the solar wind particles.
   
We are grateful to Professor J.V. Narlikar, Director for his kind hopitality,  
encouragement and all possible help during our stay at IUCAA, Pune. 
One of us (U.N.)
is grateful to Meerut College authorities for granting duty leave during the 
course of this work. The help of Dr. Sudhanshu in computation and related 
problems is thankfully acknowledged. Thanks are due to Dr. Nagendra Kumar 
for carefully reading the manuscript.

\clearpage

\begin{figure}
\includegraphics[width=\textwidth]{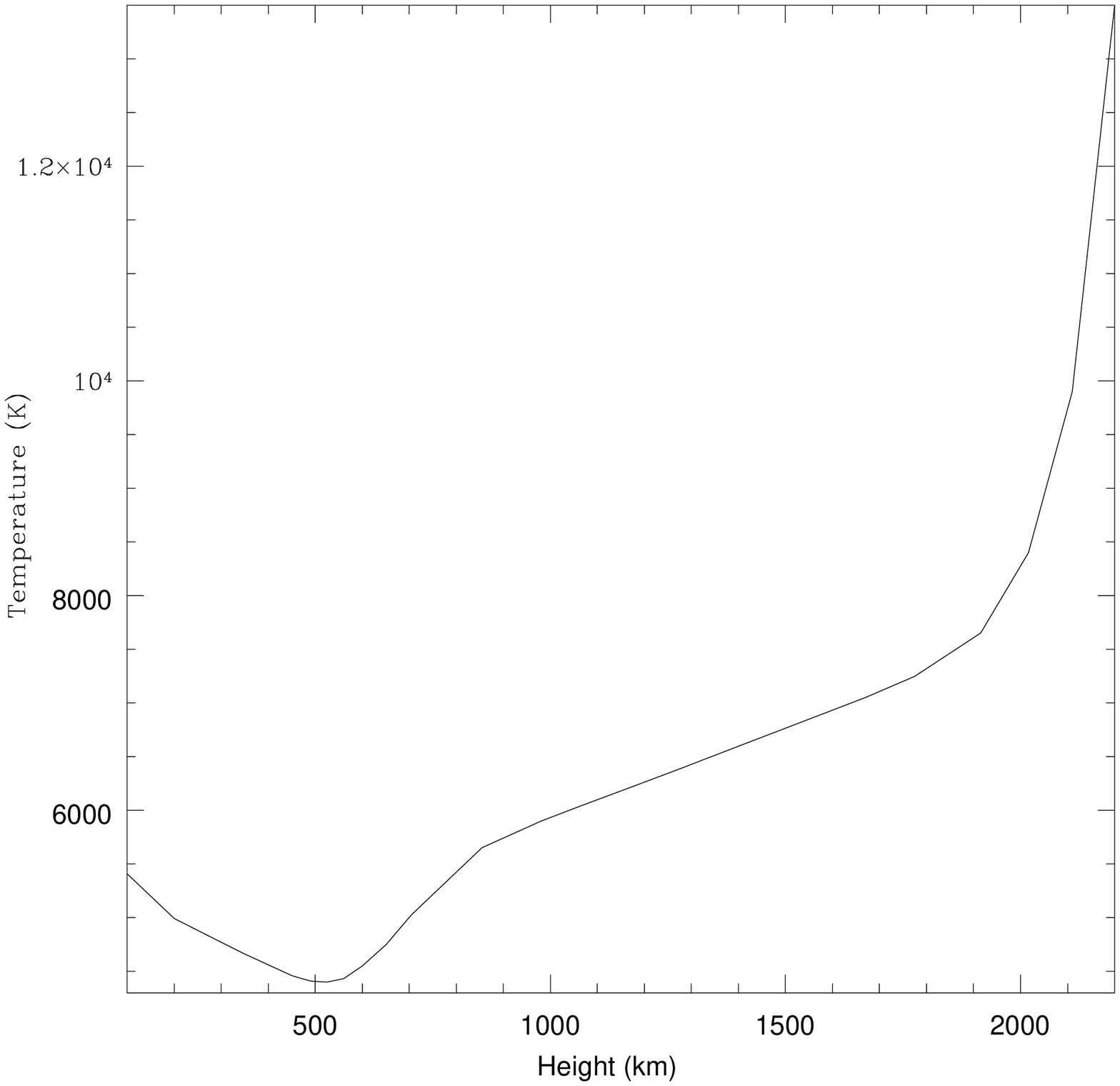}
\caption{Variation of temperature with height}
\label{fig2}
\end{figure}

\begin{figure}
\includegraphics[width=\textwidth]{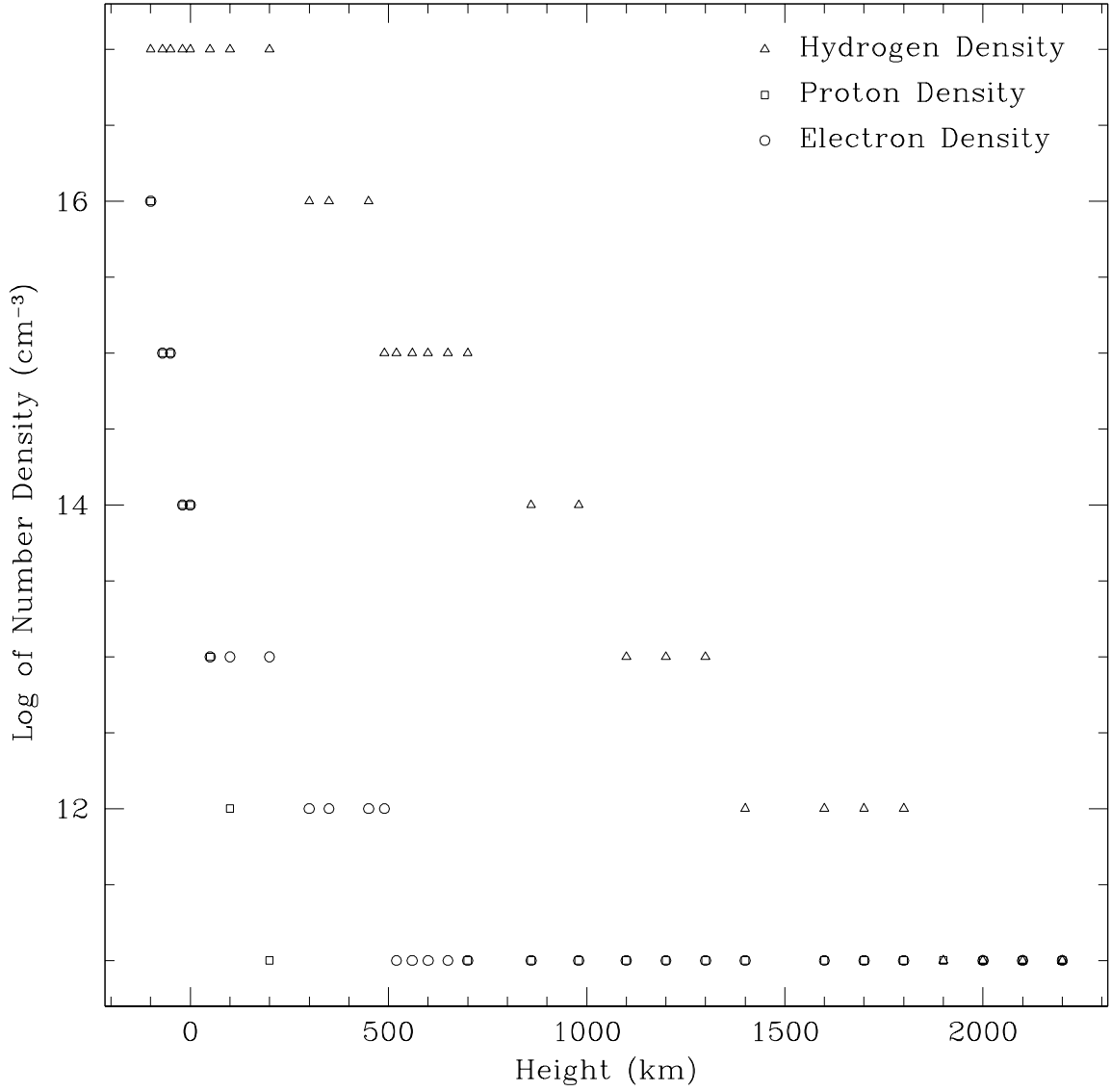}
\caption{Variation of electron, proton and hydrogen density with height}
\label{fig3}
\end{figure}

\begin{figure}
\includegraphics[width=\textwidth]{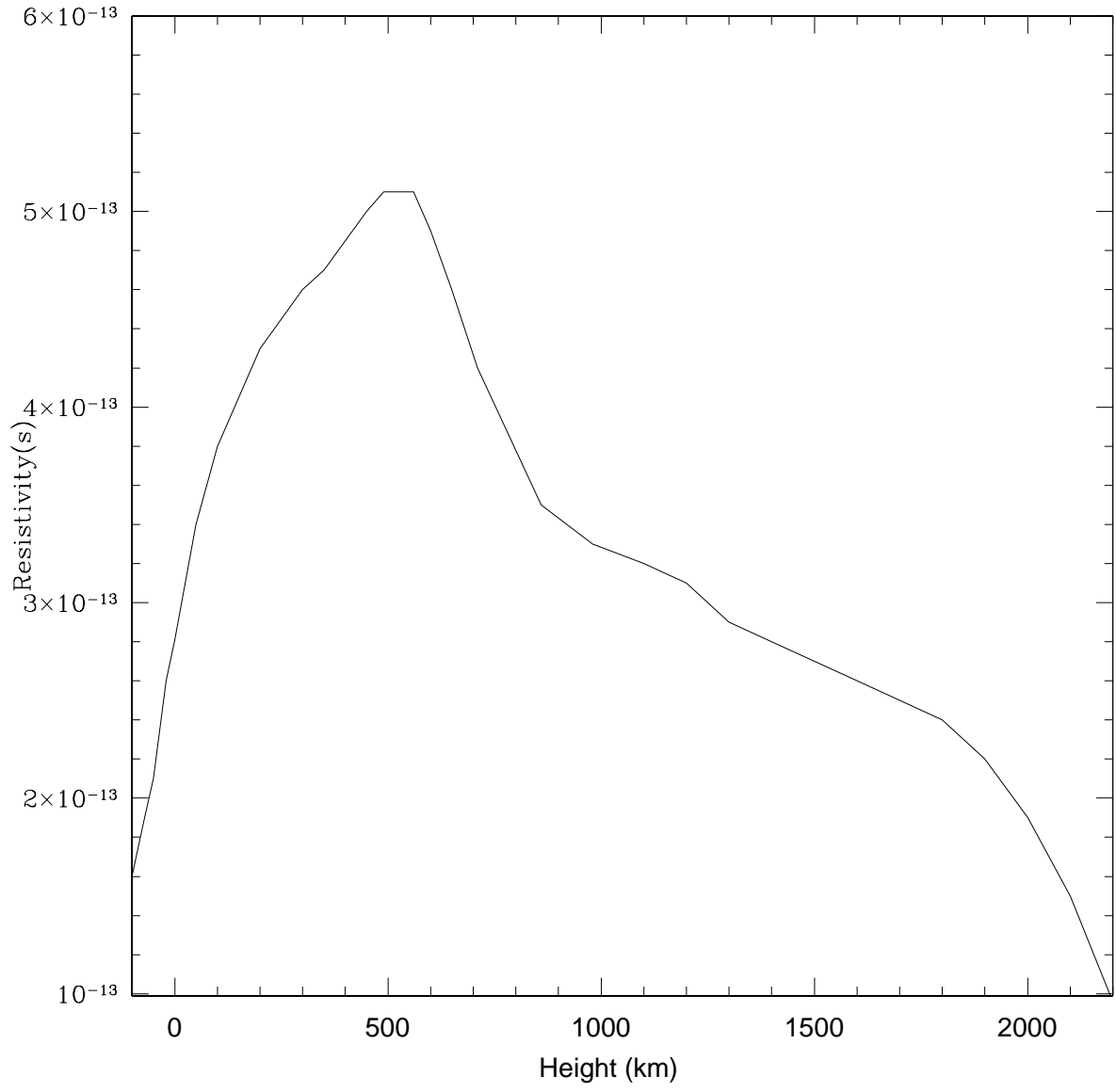}
\caption{Variation of resistivity $\eta$ with height}
\label{fig4}
\end{figure}

\begin{figure}
\includegraphics[width=\textwidth]{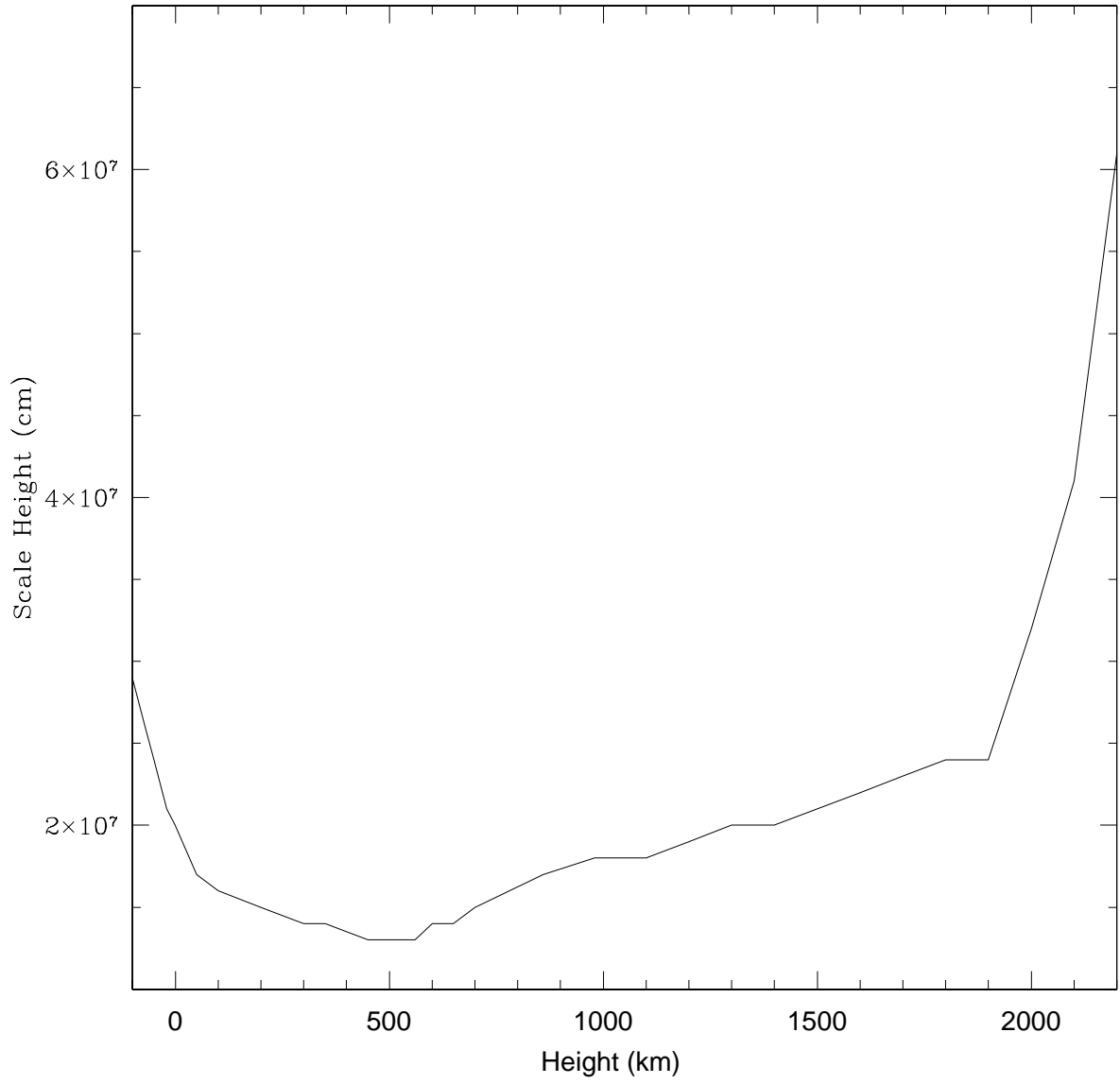}
\caption{Variation of temperature scale height H with height h}
\label{fig5}
\end{figure}

\begin{figure}
\includegraphics[width=\textwidth]{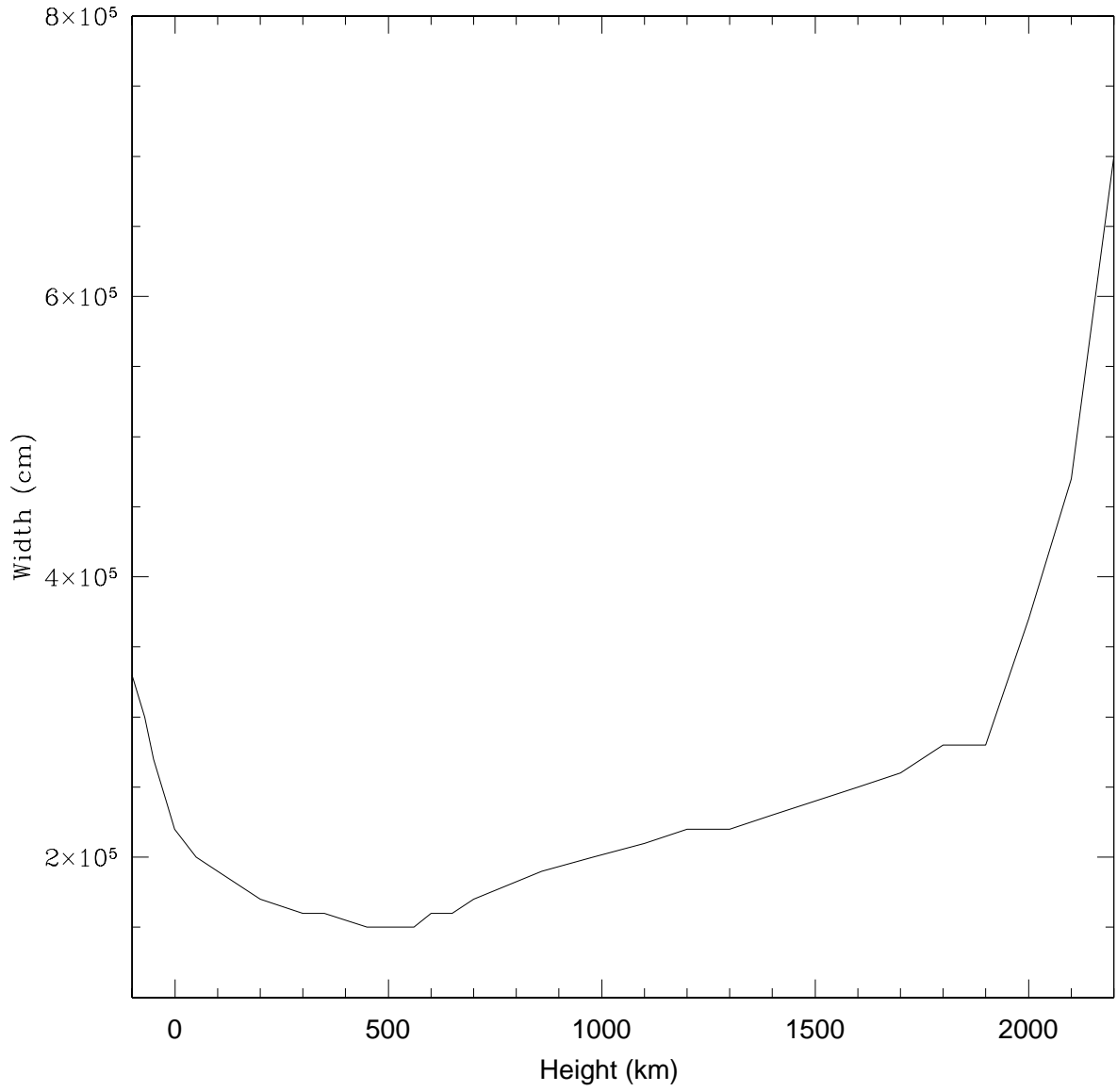}
\caption{Variation of width of reconnecting region with height}
\label{fig6}
\end{figure}

\begin{figure}
\includegraphics[width=\textwidth]{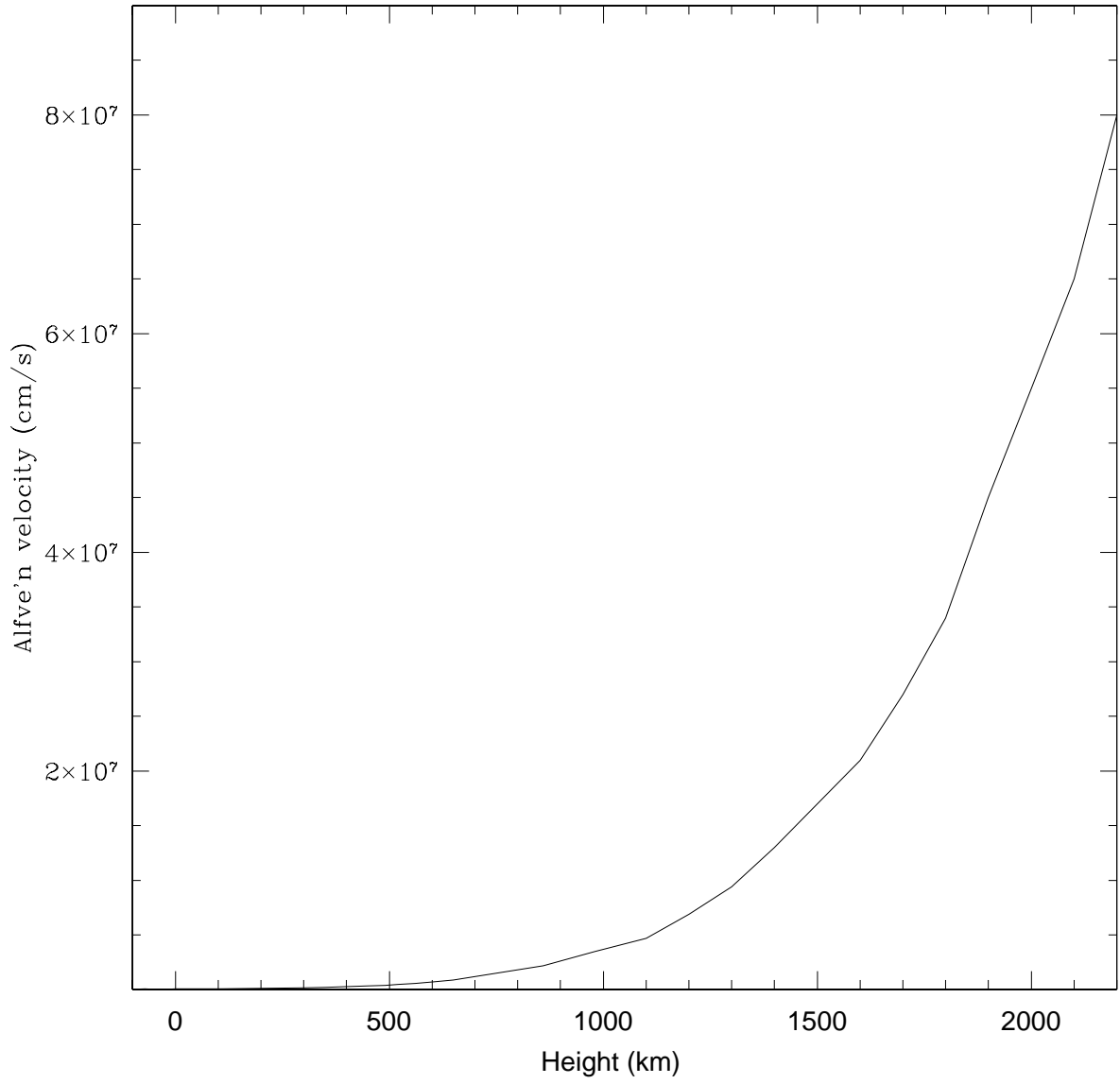}
\caption{Variation of Alfv\'en velocity with height}
\label{fig7}
\end{figure}

\begin{figure}
\includegraphics[width=\textwidth]{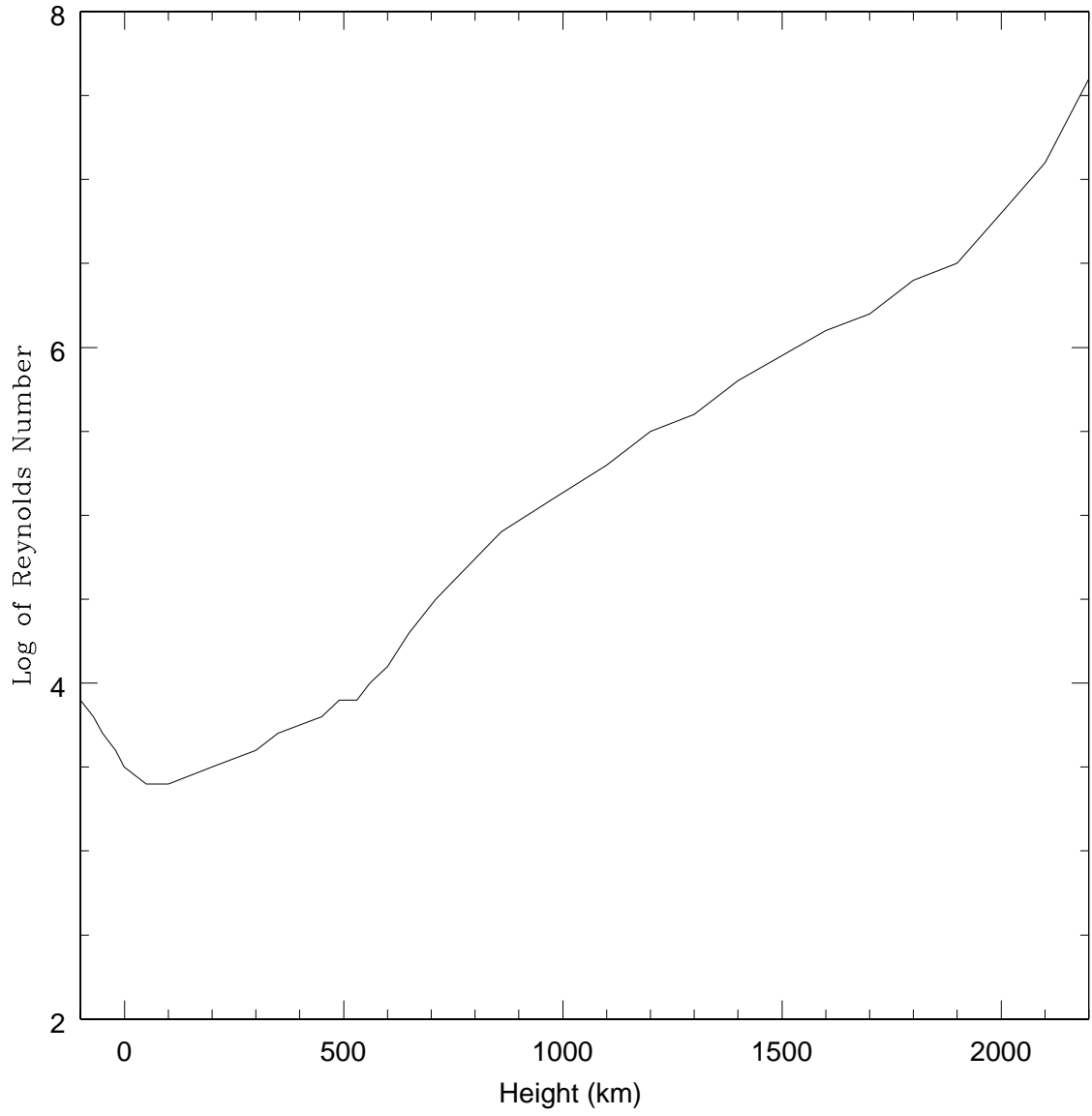}
\caption{Variation of magnetic Reynolds number with height}
\label{fig8}
\end{figure}

\begin{figure}
\includegraphics[width=\textwidth]{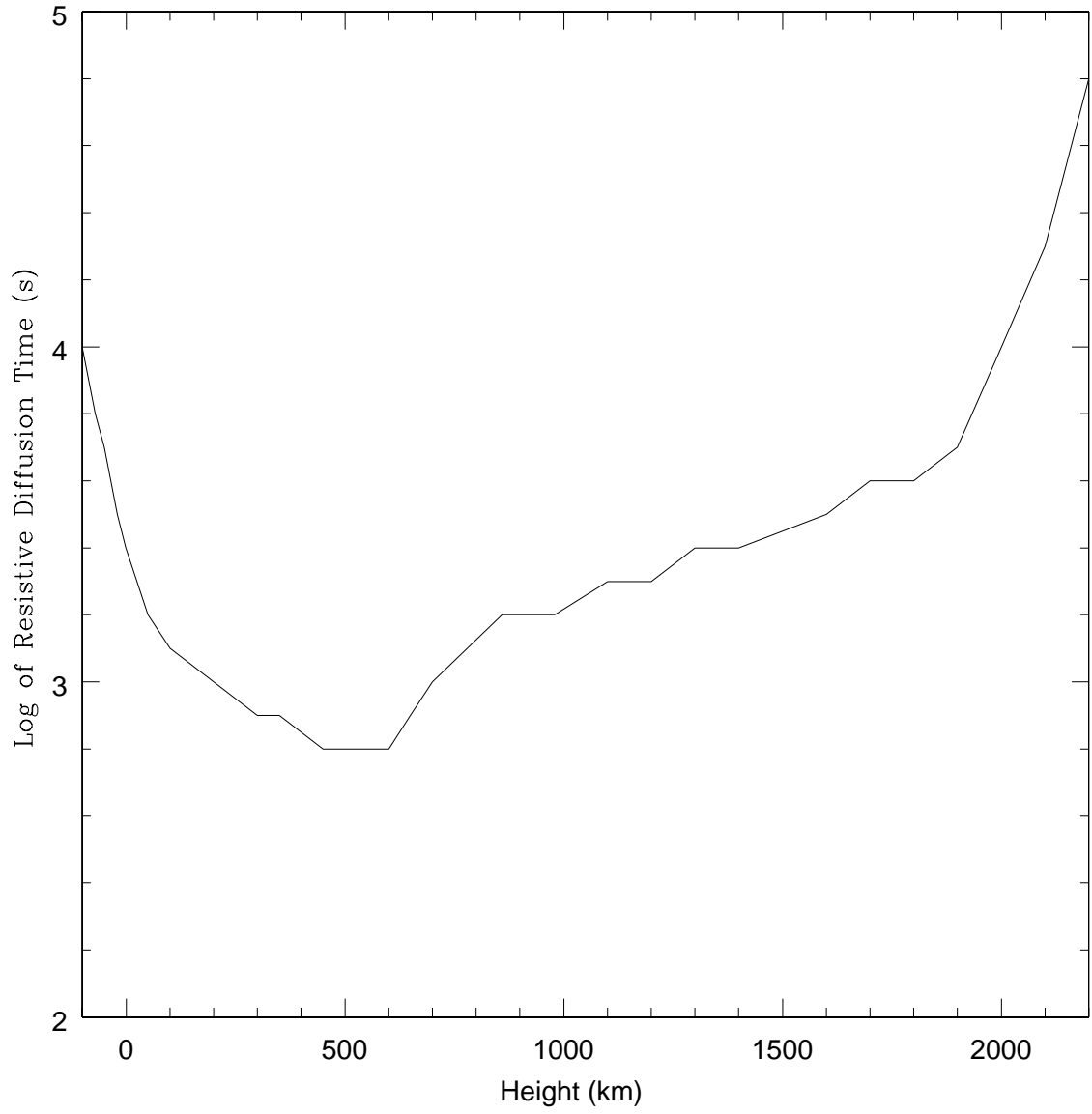}
\caption{Variation of resistive diffusion time with height}
\label{fig9}
\end{figure}

\begin{figure}
\includegraphics[width=\textwidth]{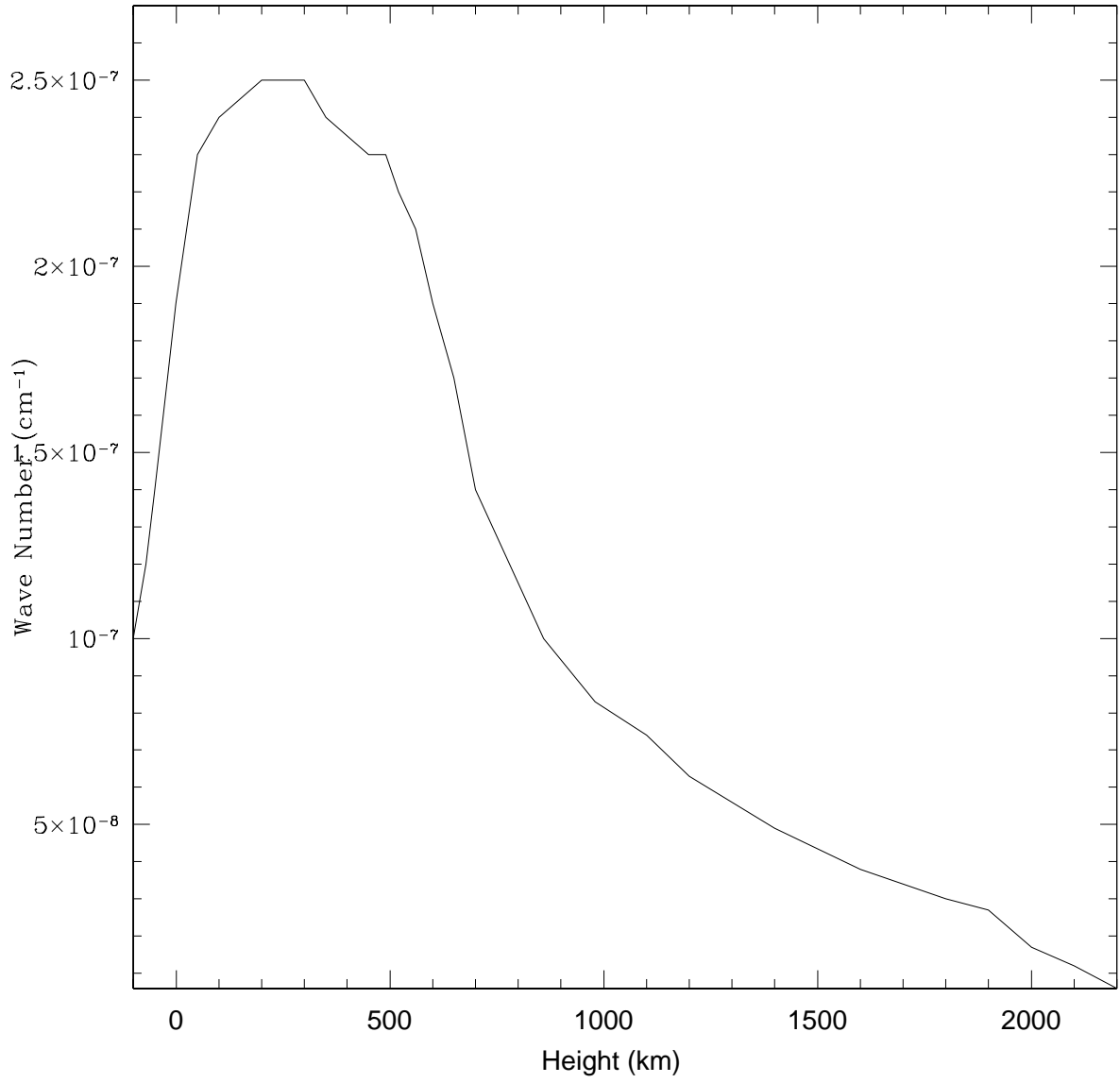}
\caption{Variation of wave number of the most rapidly growing mode with height}
\label{fig10}
\end{figure}

\begin{figure}
\includegraphics[width=\textwidth]{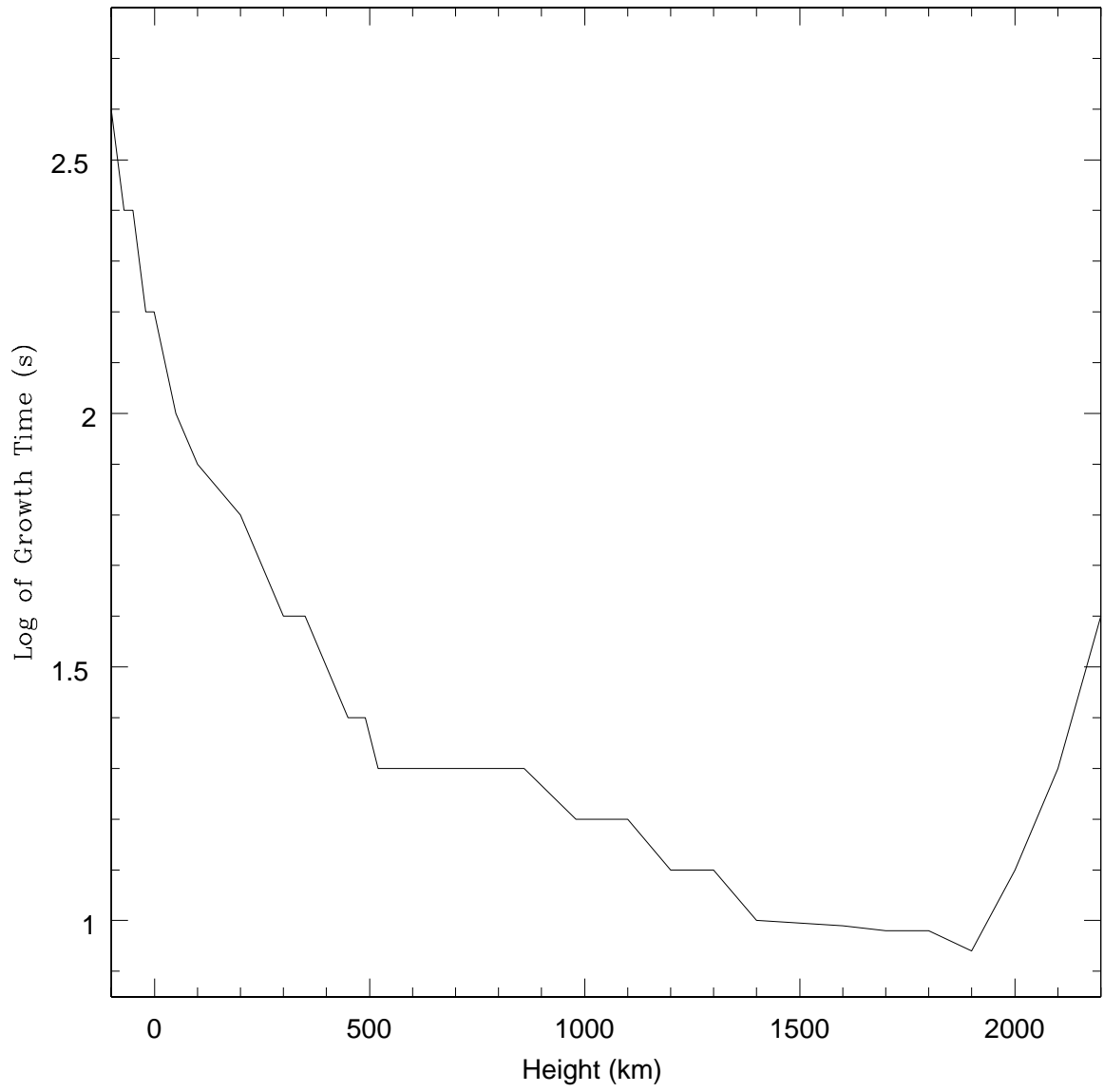}
\caption{Variation of tearing mode growth time with height}
\label{fig11}
\end{figure}

\begin{figure}
\includegraphics[width=\textwidth]{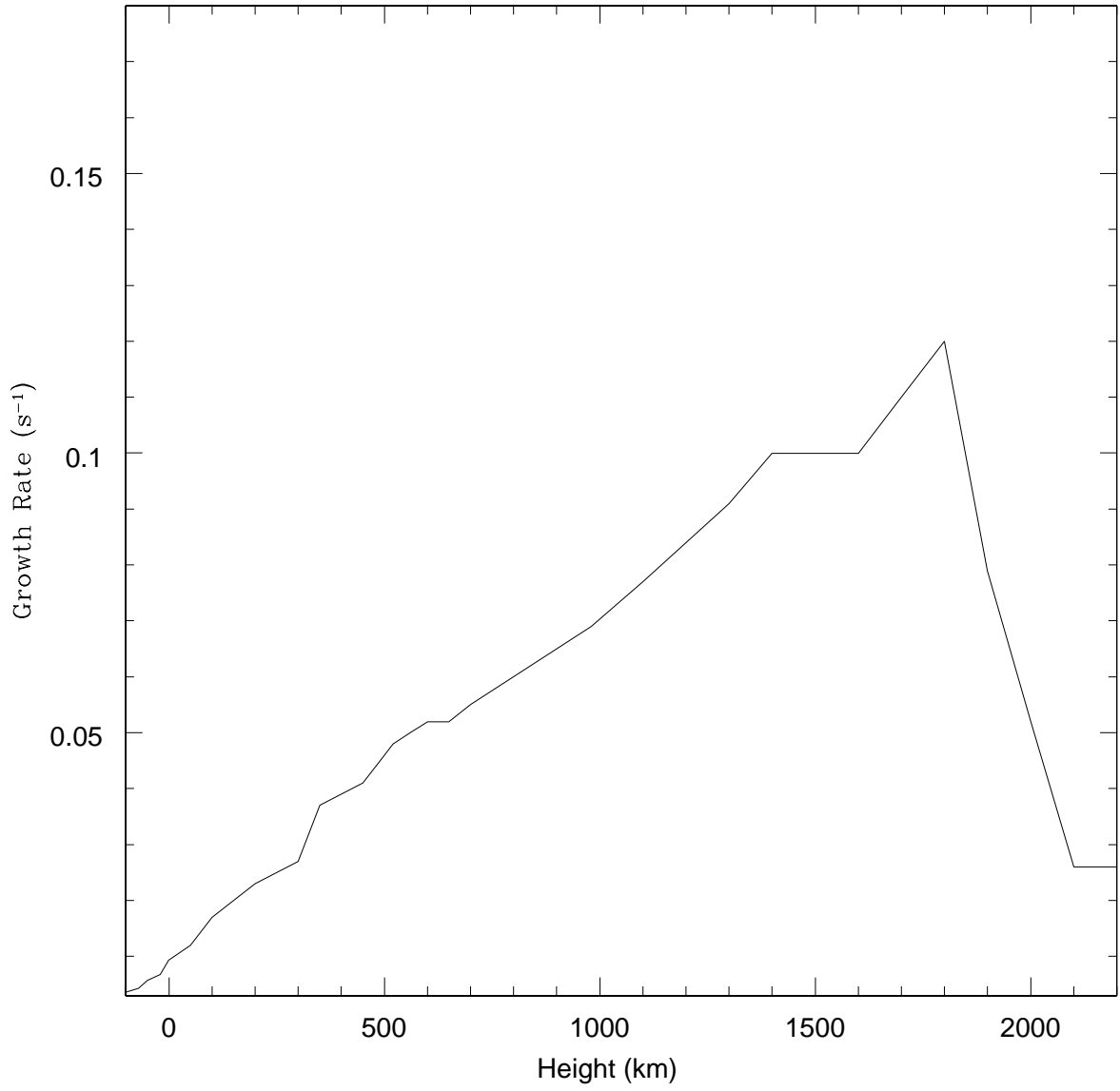}
\caption{Variation of growth rate of the tearing mode with height}
\label{fig12}
\end{figure}

\begin{figure}
\includegraphics[width=\textwidth]{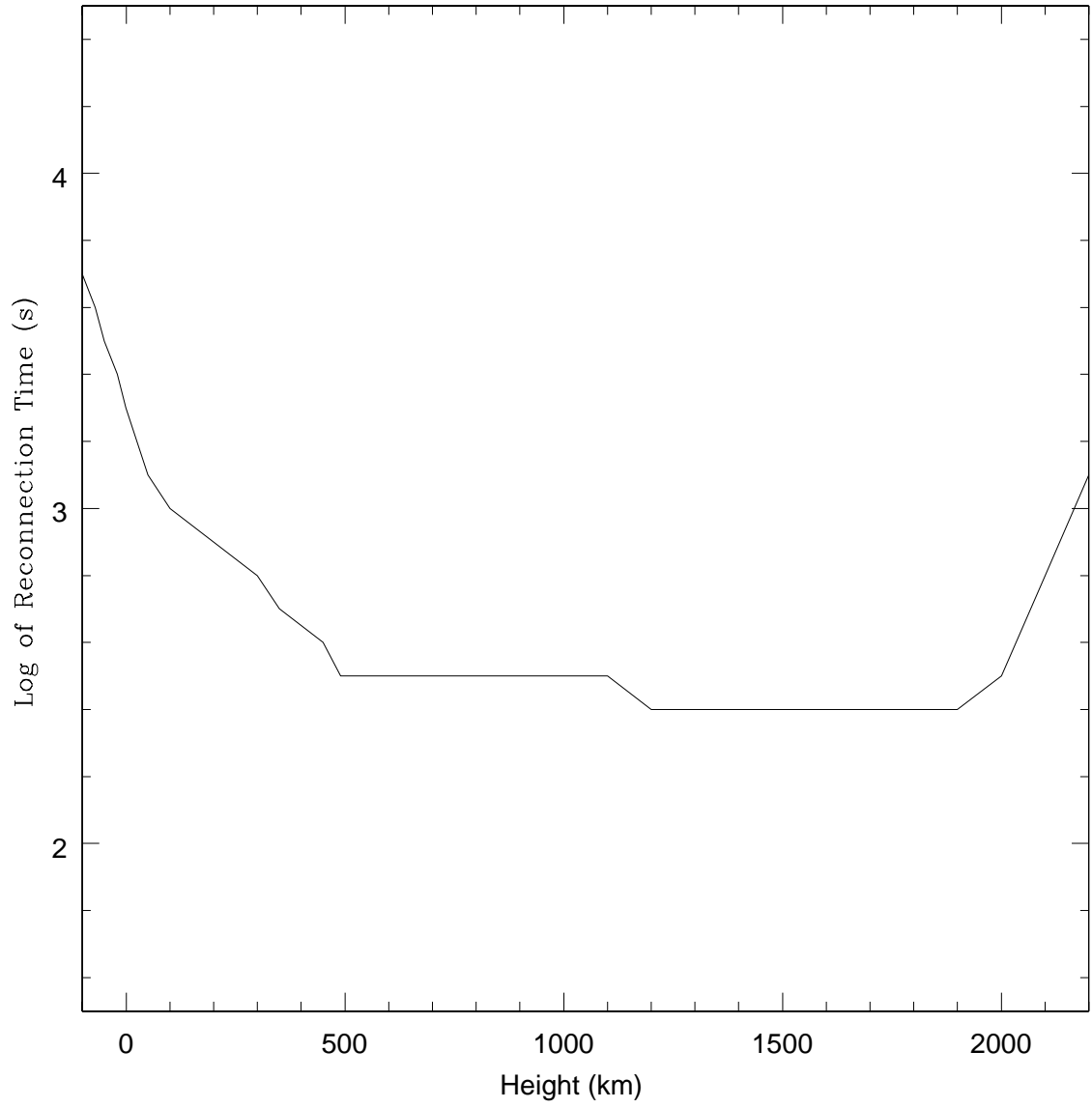}
\caption{Variation of magnetic reconnection time with height}
\label{fig13}
\end{figure}

\begin{figure}
\includegraphics[width=\textwidth]{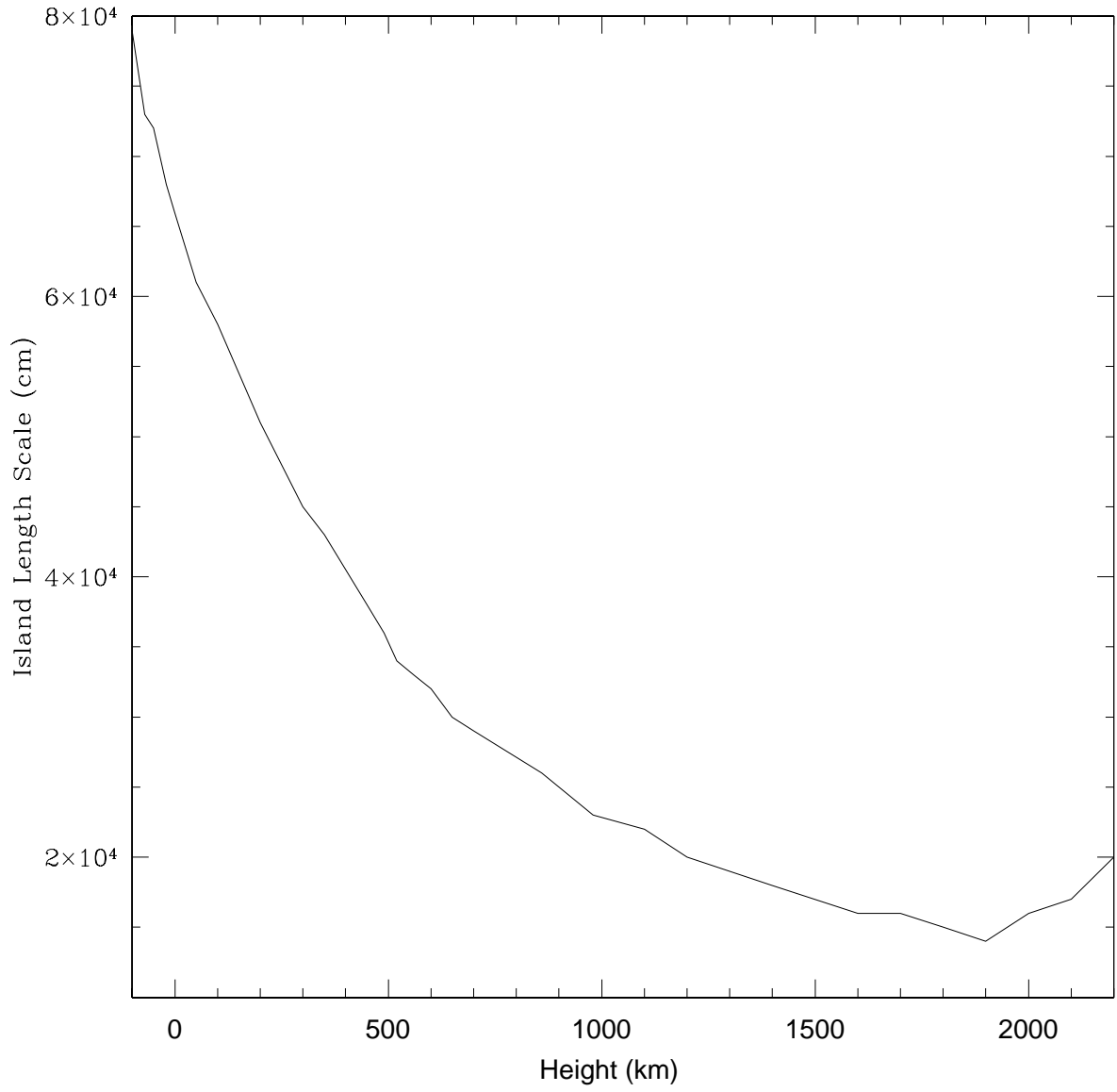}
\caption{Variation of tearing mode scale length with height}
\label{fig14}
\end{figure}

\begin{figure}
\includegraphics[width=\textwidth]{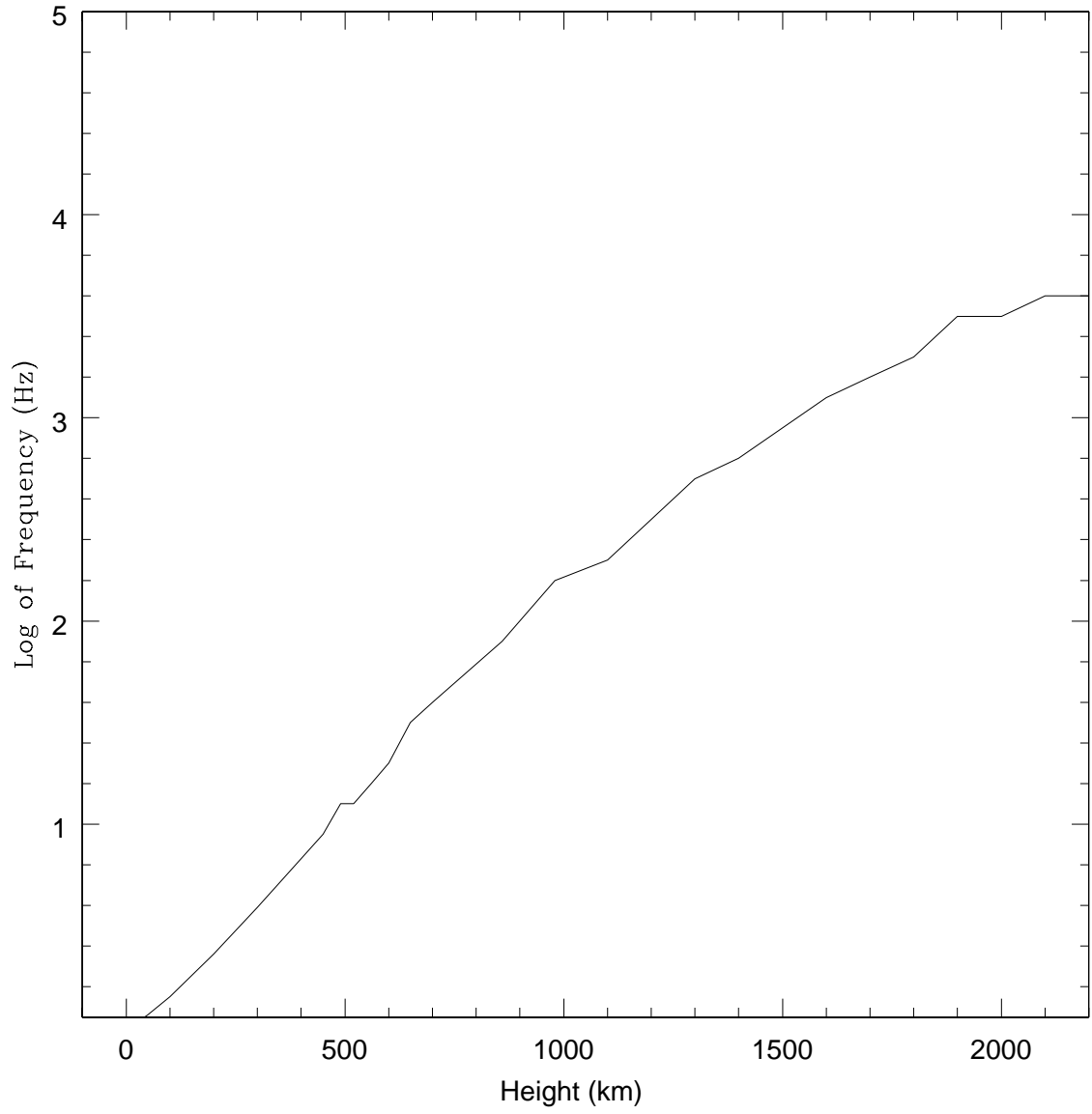}
\caption{Variation of fluctuation frequency with height}
\label{fig15}
\end{figure}

\begin{figure}
\includegraphics[width=\textwidth]{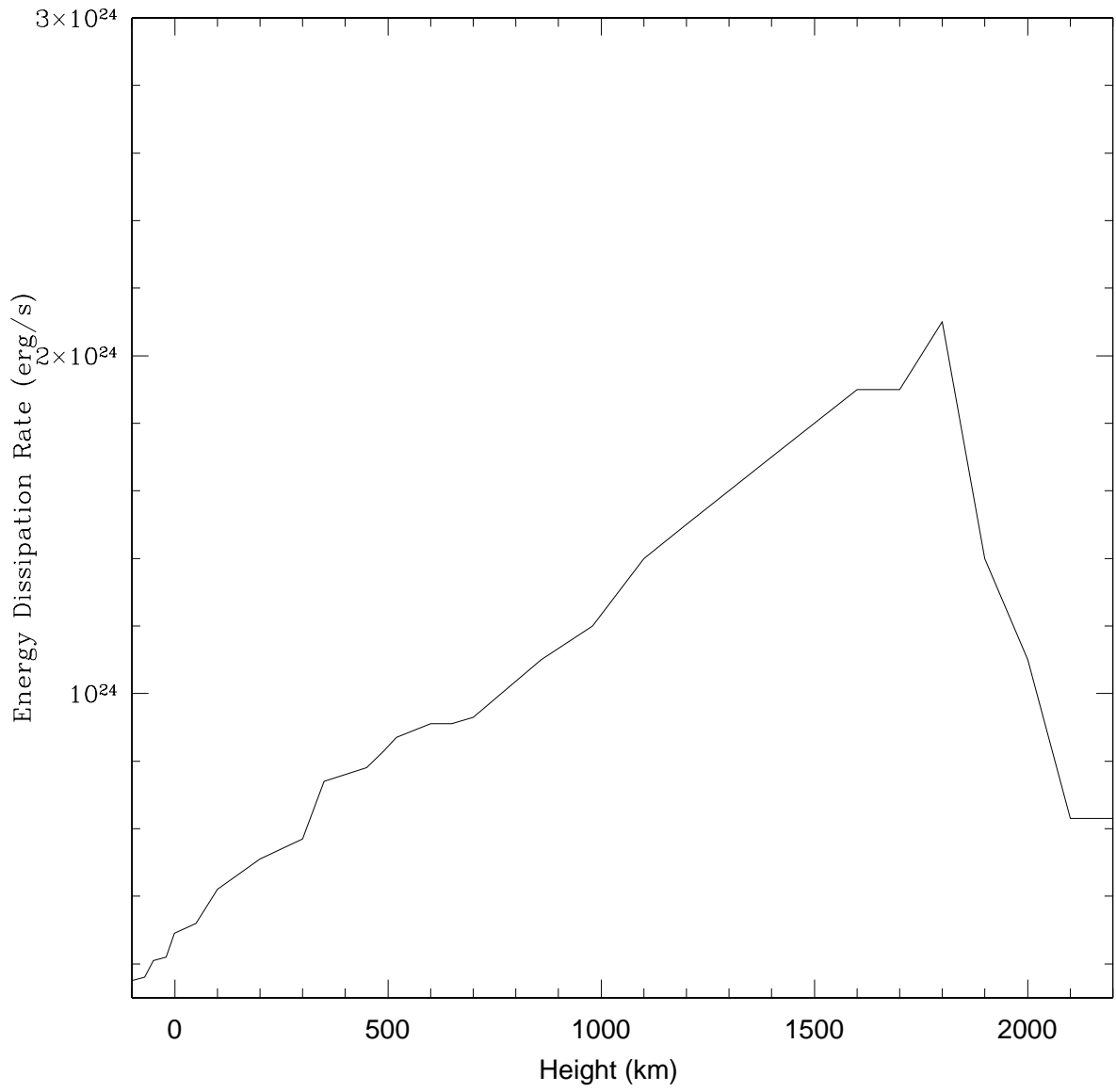}
\caption{Variation of energy dissipation rate with height}
\label{fig16}
\end{figure}

\begin{figure}
\includegraphics[width=\textwidth]{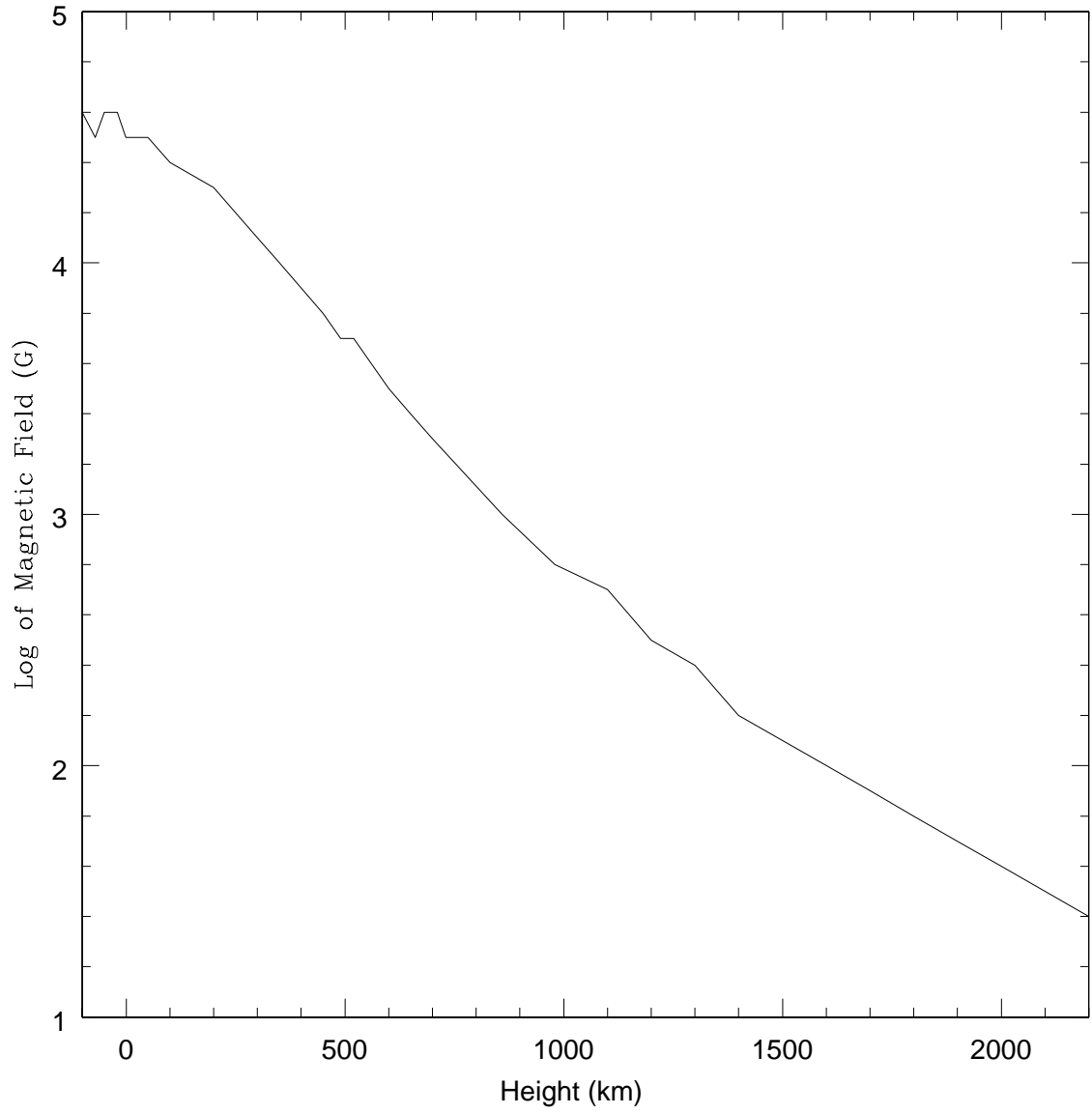}
\caption{Variation of required minimum magnetic field strength with height}
\label{fig17}
\end{figure}

\end{document}